\documentclass[aps,twocolumn,nofootinbib, preprintnumbers, superscriptaddress]{revtex4}

\usepackage{amsmath, amssymb, slashed, braket, bm}
\usepackage{graphicx}
\usepackage{epstopdf}
\usepackage{float,appendix}
\usepackage[colorlinks=true,
            linkcolor=blue,
            urlcolor=blue,
            citecolor=green,          
            bookmarks=true,
            bookmarksnumbered=true,
            breaklinks=true,
            pdfpagemode=FullScreen,
            pdfstartview=FitBH]{hyperref}
\usepackage{esint}
\usepackage[normalem]{ulem}
\usepackage{siunitx}
\usepackage{multirow}
\usepackage{xcolor}

\graphicspath{{figs/}} 

\usepackage{xcolor}
\definecolor{gesfpurple}{rgb}{0.47,0.19,0.42}

\definecolor{gesflanse}{rgb}{0.00,0.50,0.50}

\definecolor{gesfblue}{rgb}{0.08,0.42,0.76}

\definecolor{gesfred}{rgb}{1,0,0}

\definecolor{gesfwhite}{rgb}{1,1,1}

\definecolor{gesfblack}{rgb}{0,0,0}

\newcommand{\geqn}[1]{Eq.\,\hypersetup{linkcolor=blue}(\ref{#1})\hypersetup{linkcolor=blue}}
\newcommand{\gfig}[1]{{\hypersetup{linkcolor=violet}Fig.\,\ref{#1}\hypersetup{linkcolor=blue}}}
\newcommand{\gtab}[1]{{\hypersetup{linkcolor=gesflanse}Tab.\,\ref{#1}\hypersetup{linkcolor=blue}}}

\begin{document}

\title{
Coherence from Randomness: \\
\vspace{0.15cm}
Sub-keV Dark Matter Scattering off Random, Heterogeneous Materials
}

\author{Zhi-Han Liu} 
\affiliation{College of Science, China University of Petroleum (East China), Qingdao 266580, China}

\author{Shigeki Matsumoto} 
\email{shigeki.matsumoto@ipmu.jp}
\affiliation{Kavli IPMU (WPI), UTIAS, University of Tokyo, Kashiwa, 277-8583, Japan}

\author{Jie Sheng}
\email{jie.sheng@ipmu.jp}
\affiliation{Kavli IPMU (WPI), UTIAS, University of Tokyo, Kashiwa, 277-8583, Japan}

\author{Chuan-Yang Xing}
\email{cyxing@upc.edu.cn}
\affiliation{College of Science, China University of Petroleum (East China), Qingdao 266580, China}

\begin{abstract}

The sub-keV mass range has long posed a challenge for the direct detection of dark matter via elastic scattering. In this Letter, we propose a new mechanism in which dark matter, assumed to be quadratically coupled to SM particles, scatters from random heterogeneous materials with intrinsic density fluctuations, yielding an enhanced coherent response. This effect can substantially increase the total scattering rate and induce measurable accelerations of the target. Using this idea, we derive new constraints from the MICROSCOPE mission that extend into previously unexplored parameter space for sub-keV dark matter, probing cross sections down to $\sim 4\times10^{-38}\,\mathrm{cm^2}$.

\end{abstract}

\maketitle 

\section{Introduction} 

Dark matter (DM), accounting for $\sim 85\%$ of the matter content of the Universe~\cite{Planck:2018vyg}, remains one of the most profound mysteries in modern physics, with its fundamental properties still largely unknown~\cite{Bertone:2004pz, Bertone:2016nfn, Bauer:2017qwy}. Among the various strategies proposed to unveil its interactions with the Standard Model (SM), direct detection experiments play a central role~\cite{Schumann:2019eaa, Billard:2021uyg}. In particular, elastic scattering between DM and SM particles provides a theoretically well-motivated and experimentally accessible avenue, and constitutes the primary target signal in many ongoing and forthcoming direct-detection searches~\cite{CDEX:2022kcd, SENSEI:2023zdf, DarkSide-20k:2024yfq, SuperCDMS:2024yiv, PandaX:2024qfu, LZ:2024zvo, XENON:2025vwd, DAMIC-M:2025luv}.

Direct detection is further enriched by the phenomenon of coherent enhancement~\cite{Fukuda:2018omk, Akhmedov:2018wlf, Shergold:2021evs, Afek:2021vjy, Fukuda:2021drn}, which can substantially and significantly amplify the scattering rate when the DM de~Broglie wavelength, $\lambda_\chi \simeq |{\bf q}|^{-1}$, exceeds the characteristic size of the target. Here, $|{\bf q}|$ denotes the momentum transfer. In this regime, the scattering amplitudes from individual constituents add constructively, yielding a total cross section that remarkably scales with the square of the number of scatterers~\cite{Goodman:1984dc, Drukier:1986tm, Lewin:1995rx}. A well-known example occurs for weakly interacting massive particles (WIMPs) heavier than the GeV scale: because their de~Broglie wavelength exceeds the nuclear radius, they scatter coherently off all nucleons in the nucleus, resulting in an $A^{2}$ enhancement, where $A$ is the atomic mass number~\cite{Lewin:1995rx}. Analogous coherence effects for MeV-scale DM interacting with the collective electron cloud have likewise been explored to boost the scattering rate~\cite{Guo:2021imc}.

The role of coherence becomes even more striking for ultralight bosonic DM that is quadratically coupled to SM fields~\cite{Fukuda:2018omk, Afek:2021vjy, Day:2023mkb, Luo:2024ocg, Gan:2025nlu, Matsumoto:2025rcz, Acevedo:2025rqu}. In such scenarios, the de Broglie wavelength can easily exceed macroscopic scales, for example, meV-scale DM has $\lambda_\chi$ of order meters, allowing it to scatter coherently off an entire bulk target. This macroscopic coherence leads to an enormous enhancement of the total scattering cross section proportional to $N_{\mathrm{tot}}^{2}$, where $N_{\mathrm{tot}} \sim 10^{23}$ denotes the total number of nucleons in a laboratory-scale detector~\cite{Day:2023mkb, Luo:2024ocg, Matsumoto:2025rcz}. Despite this dramatic amplification, the energy transferred in a single scattering event is exceedingly small, typically $\Delta E \lesssim 10^{-5}\,\mathrm{eV}$, rendering calorimetric detection infeasible. Importantly, while the energy transfer is suppressed by the large target mass, the momentum transfer is not; the cumulative impulse from the high event rate can therefore induce a tiny but potentially measurable acceleration of the target~\cite{Day:2023mkb, Luo:2024ocg, Matsumoto:2025rcz}. This observation motivates force- or acceleration-based search strategies, such as asymmetric precision torsion-balance experiments, exploiting macroscopic coherence to probe quadratically coupled DM in the sub-eV mass regime~\cite{Luo:2024ocg, Matsumoto:2025rcz}.

As the DM mass increases, its de~Broglie wavelength becomes shorter than the detector size $R$, and macroscopic coherence is lost, since scattering amplitudes from distant regions acquire varying phases and interfere destructively. For a spherical, homogeneous target, the resulting form-factor suppression scales as $(|{\bf q}|\,R)^{-4}$~\cite{Afek:2021vjy, Matsumoto:2025rcz}, so the effective cross section decreases rapidly as the momentum transfer, thus the DM mass, grows. In the sub-keV range, the coherence length reaches only micrometer scales, far smaller than typical force sensors, and the energy deposited in scattering, $\Delta E \sim 10^{-3}\,\mathrm{eV}$, is still below calorimetric thresholds~\cite{Essig:2022dfa, Du:2022dxf, Das:2023cbv, Essig:2024wtj, Helis:2024vhr, Hochberg:2025dom}. This combination of suppressed coherence and tiny energy deposition makes this sub-keV mass regime difficult to probe with existing direct-detection methods~\cite{Baker:2023kwz, Das:2023cbv, Hochberg:2025dom}.

In this work, we show that the loss of macroscopic coherence at sub-keV masses does not prevent significant enhancement: even when the DM wavelength $\lambda_\chi$ is much smaller than the detector size, coherent scattering can still occur locally if the target is a random, heterogeneous material. Such materials contain many micron-scale patches with different densities or compositions, and when the patch size $\xi$ is comparable to the DM wavelength, $\xi \sim |{\bf q}|^{-1}$, DM scatters coherently within each patch, while contributions from distant patches add only incoherently. With a nucleon density of the detector $n_0 \sim N_{\rm tot}/R^3$, the total cross section scales as $\sigma_{\rm tot} \propto [N_{\rm tot}/(n_0 \xi^3)] (n_0 \xi^3)^2 \sim N_{\rm tot}\,n_0\,\xi^3 \sim N_{\rm tot}^2 (|{\bf q}|R)^{-3}$, giving a parametric enhancement over the homogeneous case. Using a general formalism to quantify this effect, we show that heterogeneous materials can substantially boost the scattering rate of quadratically coupled DM with sub-keV masses. Applying the framework to the MICROSCOPE test masses, we obtain direct-detection bounds on the DM–nucleon cross section as strong as $10^{-38}\,\mathrm{cm}^2$, and argue that future torsion-balance experiments using materials with larger density contrasts could improve the sensitivity by additional orders of magnitude, potentially reaching scattering cross sections of order $10^{-42}\,\mathrm{cm}^2$.

\section{Coherence in Randomness}

Considering a quadratic coupling between DM and nucleons, such as an interaction of the form $(m_N/\Lambda^2) \chi^2 \bar{N} N$ for scalar\footnote{
    In the mass range below $\sim 0.2\,\mathrm{keV}$, DM must be bosonic due to the so-called Tremaine--Gunn bound\,\cite{Tremaine:1979we, Domcke:2014kla, Alvey:2020xsk}, whereas in the mass interval of $0.2\text{--}1\,\mathrm{keV}$ DM can be either bosonic or fermionic.}
DM $\chi$, where $N = n,p$ denotes neutrons or protons, the DM can scatter off terrestrial targets. In the sub-keV mass regime, the DM de Broglie wavelength $(m_\chi v_\chi)^{-1}$, with a typical velocity of the DM $v_\chi \sim 10^{-3}$ in the laboratory frame, exceeds the micron scale and is much larger than the interatomic spacing of order $10^{-10}\,\mathrm{m}$. Therefore, one can adopt a mean-field description of the target, in which the discrete atomic structure is replaced by a smooth number-density profile $n(\mathbf{r})$.

Given the weakness of the DM interaction, the Born approximation can be employed. The differential scattered wave from all nucleons within a volume element,
\begin{equation}
    d\Psi_{\rm sc}(\mathbf{r}) =
    f(\theta)\, e^{i\mathbf{q}\cdot\mathbf{r}_1}\,
    \frac{e^{ik|{\bf r}|}}{|{\bf r}|}\,
    n(\mathbf{r}_1)\, d^3\mathbf{r}_1 ,
\end{equation}
is obtained by summing the single-nucleon amplitudes $f(\theta)$, each weighted by the phase factor $\mathbf{q}\cdot\mathbf{r}_1$ with $\mathbf{q}=\mathbf{k}-\mathbf{k}'$, assuming a spin-independent interaction between the DM and a nucleon. Here, $\theta$ is the scattering angle, and $k = |\mathbf{k}| = |\mathbf{k}'|$ denotes the DM wavenumber. Since the DM is much lighter than the nucleons, the magnitudes of the initial and final momenta, $\mathbf{k}$ and $\mathbf{k}'$, remain essentially unchanged during scattering. Integrating over the entire target and squaring the resulting amplitude gives the total differential cross section.
\begin{equation}
    \frac{d\sigma_{\rm tot}}{d\Omega}
    =
    \frac{d\sigma_{\chi N}}{d\Omega}
    \int d^3\mathbf{r}_1 d^3\mathbf{r}_2\;
    e^{i \mathbf{q}\cdot (\mathbf{r}_2-\mathbf{r}_1)}\,
    n(\mathbf{r}_1)\,n(\mathbf{r}_2),
    \label{eq:cross_section_general}
\end{equation}
where $d\sigma_{\chi N}/d\Omega = |f(\theta)|^2$ is the differential cross section for scattering between the DM particle and a nucleon.

For a homogeneous target of volume $V$ with constant density $n(\mathbf{r}) = n_0$, the double integral in Eq.~\eqref{eq:cross_section_general} depends only on the target geometry and overall size. With the total number of scatterers given by $N_{\mathrm{tot}} = n_0 V$, the differential cross section can then be factorized as
\begin{equation}
    \frac{d\sigma_{\rm tot}}{d\Omega}
    =
    N_\mathrm{tot}^2\,|F_{\rm homo}(\mathbf{q})|^2\,
    \frac{d\sigma_{\chi N}}{d\Omega},
    \label{cross_section_factorized}
\end{equation}
with the dimensionless form factor defined as
\begin{equation}
    |F_{\rm homo}(\mathbf{q})|^2 \equiv
    \frac{n_0^2}{N_\mathrm{tot}^2}
    \int d^3\mathbf{r}_1 d^3\mathbf{r}_2\;
    e^{i \mathbf{q}\cdot (\mathbf{r}_2-\mathbf{r}_1)}.
    \label{form_factor_homogeneous}
\end{equation}
If the separation $|{\bf r}_2 - {\bf r}_1|$, which is typically of the order of the target size, is much smaller than the DM wavelength, $|{\bf r}_2 - {\bf r}_1| \ll q^{-1} \equiv |\mathbf{q}|^{-1}$, then $|F_{\rm homo}| \simeq 1$, i.e., the scattering is coherent across the entire target, yielding an $N_{\mathrm{tot}}^{\,2}$ enhancement relative to the single-nucleon scattering cross section. Conversely, when the wavelength becomes shorter, the phase oscillates rapidly, the integral is suppressed, and coherence is consequently lost.

For a homogeneous sphere of radius $R$, the form factor can be analytically evaluated in the following form:
\begin{equation}
    |F_{\rm homo}(\mathbf{q})|^2
    =
    \left[\frac{3\big(\sin(qR)-qR\cos(qR)\big)}{(qR)^3}\right]^2,
    \label{form_factor_spherical_homogeneous}
\end{equation}
whose envelope scales as $(qR)^{-4}$ at large $qR$. Denoting $N_c \equiv (4\pi/3)\,\lambda_\chi^{3} n_0$ as the number of scatterers within one DM wavelength, the total cross section scaling for DM scattering off a homogeneous target can be written as $\sigma_{\rm tot} \propto N_{\rm tot}^2 |F_{\rm homo}|^2 \propto R^6 \times (\lambda_\chi^4 / R^4) \propto N_{\rm tot}^{2/3} N_c^{4/3}$, where we have used the relation $N_{\rm tot(c)} \propto R^3 (\lambda_\chi^3)$. This scaling drops rapidly as the DM wavelength becomes shorter.

We now arrive at the key point of our work: the total scattering cross section of sub-keV DM can be greatly enhanced when the target is a random heterogeneous material. In a random medium, the exact density function is not known deterministically and can be defined as
\begin{equation}
    n(\mathbf{r})\equiv n_0+\delta n(\mathbf{r}),
\label{nr}
\end{equation}
with $n_0 \equiv \langle n(\mathbf{r}) \rangle$ being the average density and $\delta n$ the corresponding density fluctuation, satisfying $\langle \delta n(\mathbf{r}) \rangle = 0$. Here $\langle \cdots \rangle$ denotes an ensemble average. The properties of a random material are largely encoded in its two-point correlation function $\langle n(\mathbf{r}_1)\, n(\mathbf{r}_2)\rangle$, which measures how the density at $\mathbf{r}_2$ co-varies with that at $\mathbf{r}_1$ in space. A large value of $\langle n(\mathbf{r}_1)\, n(\mathbf{r}_2)\rangle$ indicates that the material tends to exhibit similar densities at the two points. According to Eq.~\eqref{nr}, this correlation naturally splits as
\begin{equation}
    \langle n(\mathbf{r}_1) n(\mathbf{r}_2) \rangle
    = 
    n_0^2 
    + 
    \langle \delta n(\mathbf{r}_1) \delta n(\mathbf{r}_2) \rangle .
    \label{density_correlation}
\end{equation}
The first term corresponds to the contribution from the homogeneous average density, while the second term encodes the resulting effect of local fluctuations.

Similarly, the total scattering cross section for DM interacting with a target composed of a random material should also be obtained by averaging Eq.~\eqref{eq:cross_section_general}, yielding
\begin{equation}
    \left\langle \frac{d\sigma_{\rm tot}}{d\Omega} \right\rangle
    =
    N_\mathrm{tot}^2\,|F(\mathbf{q})|^2\,
    \frac{d\sigma_{\chi N}}{d\Omega} .
\end{equation}
As indicated in Eq.~\eqref{density_correlation}, the form factor in this case
\begin{equation}
    |F(\mathbf{q})|^2 \equiv |F_{\rm homo}(\mathbf{q})|^2 + |F_{\rm fluc}(\mathbf{q})|^2,
\end{equation}
receives contributions both from the form factor of a homogeneous medium, as in Eq.~\eqref{form_factor_homogeneous}, and from the effects of density fluctuations that arise in random materials,
\begin{equation}
    |F_\mathrm{fluc}(\mathbf{q})|^2 \equiv \frac{1}{N_\mathrm{tot}^2}
    \int d^3\mathbf{r}_1 d^3\mathbf{r}_2\;
    e^{i\mathbf{q}\cdot(\mathbf{r}_2-\mathbf{r}_1)}\,
    \langle \delta n(\mathbf{r}_1) \delta n(\mathbf{r}_2)\rangle,
    \label{form_factor_random}
\end{equation}
which captures the distinctive scattering properties of random heterogeneous media in a consistent manner.

Random heterogeneous materials typically exhibit clustering, in which atoms aggregate into groups whose characteristic size~$\xi$ sets the scale over which the density field is correlated. At coincident points ($\mathbf{r}_1 = \mathbf{r}_2$), the correlation of the density fluctuation $\delta n$ is maximal, with $\langle \delta n^2(\mathbf{r}) \rangle = \overline{\delta n^2}$ defining the variance. For separations satisfying $|\mathbf{r}_2 - \mathbf{r}_1| \ll \xi$, the fluctuations $\delta n(\mathbf{r}_1)$ and $\delta n(\mathbf{r}_2)$ take similar values, yielding a large correlation $\langle \delta n(\mathbf{r}_1)\,\delta n(\mathbf{r}_2)\rangle$. In contrast, when $|\mathbf{r}_2 - \mathbf{r}_1| \gg \xi$, the fluctuations become statistically independent and $\langle \delta n(\mathbf{r}_1)\,\delta n(\mathbf{r}_2)\rangle \approx \langle \delta n(\mathbf{r}_1)\rangle\,\langle \delta n(\mathbf{r}_2)\rangle = 0$. This behavior typically reflects a generic property of random media: correlations are strong at short distances and decay to zero at long distances, with the length scale~$\xi$ characterizing this decay referred to as the correlation length.

A two-phase system whose components have random shapes and sizes is often called
a Debye random medium~\cite{Debye_1949,Debye_1957}. Its correlation function
$\langle \delta n(\mathbf{r}_1)\,\delta n(\mathbf{r}_2)\rangle$ is closely
approximated by an exponential form~\cite{Jiao:2007aaa},
\begin{equation}
    \langle \delta n (\mathbf{r}_1)\,\delta n (\mathbf{r}_2) \rangle
    =
    \overline{\delta n^2}\,
    e^{- |\mathbf{r}_2 - \mathbf{r}_1| / \xi} 
    \label{correlation_function_exponential}
\end{equation}
with a strong short-distance correlation that decreases and approaches zero at sufficiently large separations.

\begin{figure}[!t]
    \centering
    \includegraphics[width=8cm]{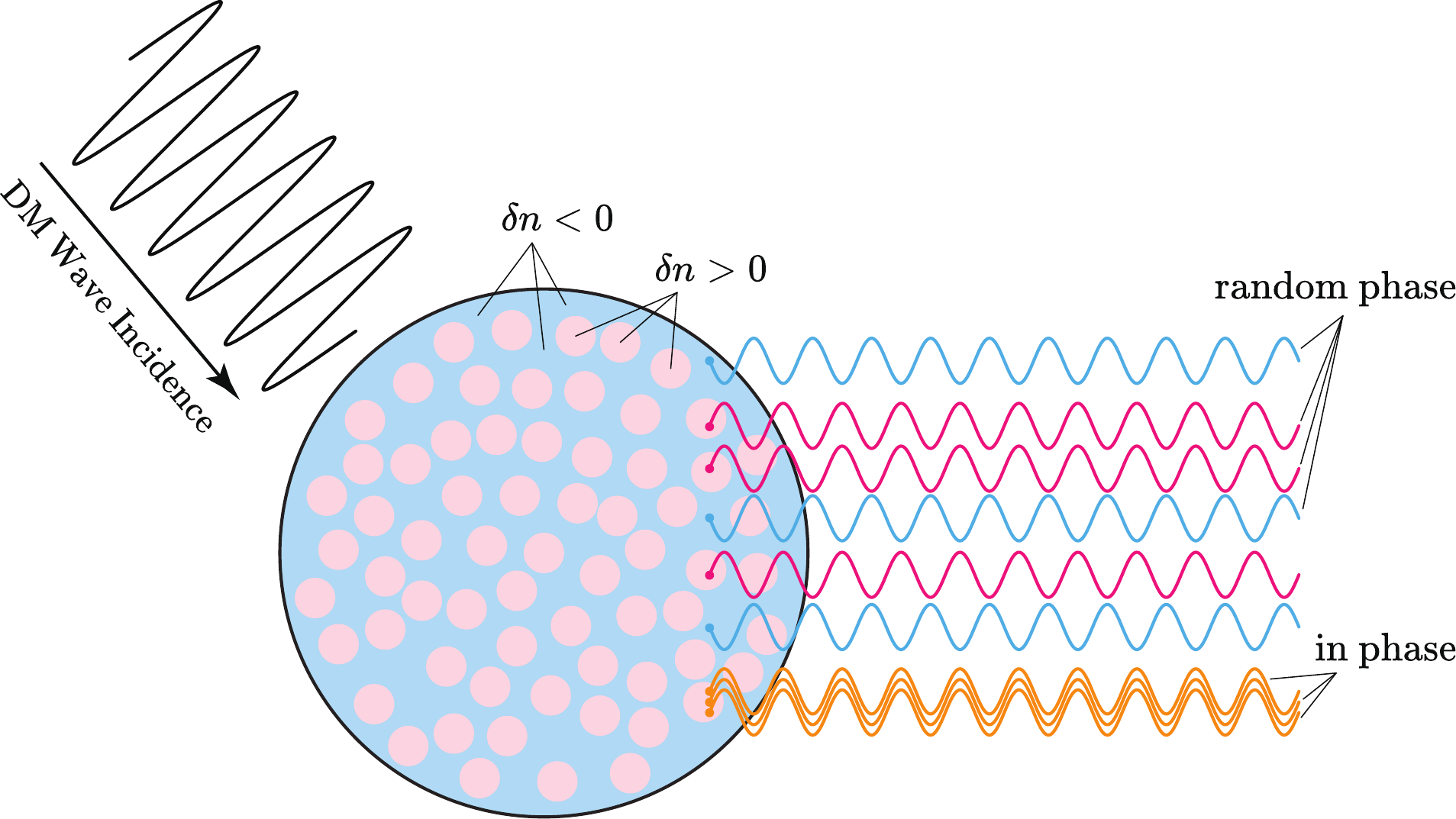}
    \caption{Scattered waves from particles within the same cluster (orange) are in phase and therefore add coherently. In contrast, waves originating from different clusters (red and blue) carry random relative phases due to the fluctuating sign of the density contrast $\delta n$, so their average contribution vanishes, and the signals from different clusters add only incoherently.}
    \label{fig:illustration}
\end{figure}

The fluctuation-induced form factor of a spherical target with radius $R$ can be obtained by inserting Eq.~\eqref{correlation_function_exponential} into Eq.~\eqref{form_factor_random} and then simply taking the limit $R \gg \xi$,
\begin{equation}
    |F_\mathrm{fluc}(\mathbf{q})|^2
    =
    \frac{\overline{\delta n^2}}{n_0^2}\,
    \frac{6 \,\xi^3}{R^3 \big(1 + q^2 \xi^2\big)^2}.
    \label{form_factor_fluctuation}
\end{equation}
For $\xi \simeq q^{-1} \sim \lambda_\chi$, this fluctuation-induced form factor scales as $\xi^{3}/R^{3} \sim (qR)^{-3}$. For materials with large density contrast, $\overline{\delta n^{2}} \simeq n_{0}^{2}$, and recalling that $|F_\mathrm{homo}(\mathbf{q})|^{2} \propto (qR)^{-4}$, the fluctuation form factor dominates in the regime $q^{-1} \ll R$, i.e., $|F_\mathrm{fluc}(\mathbf{q})|^{2} \gg |F_\mathrm{homo}(\mathbf{q})|^{2}$, so that $|F(\mathbf{q})|^{2} \approx |F_\mathrm{fluc}(\mathbf{q})|^{2}$. This implies that the total cross section scales as $\sigma_{\rm tot} \propto N_\mathrm{tot}^{2}\,|F_\mathrm{fluc}(\mathbf{q})|^{2} \propto N_\mathrm{tot}\,N_{c}$, where $N_{c} \equiv (4\pi/3)\,\lambda_\chi^{3} n_{0} \sim (4\pi/3)\,\xi^{3} n_{0}$. Thus, the total scattering cross section of sub-keV DM can receive a substantial enhancement by a factor of $(N_\mathrm{tot}/N_{c})^{1/3} \sim (R/\xi)$ compared to the case of a homogeneous material.

This realizes the intuitive picture emphasized in the Introduction: DM scatters \emph{coherently} within each cluster, whereas the contributions from distinct clusters add \emph{incoherently}, as illustrated clearly in Fig.~\ref{fig:illustration}. At fixed $q$, the form factor in Eq.~\eqref{form_factor_fluctuation} is maximized at $\xi = \sqrt{3}/q$, which reflects the optimal balance between the cluster size and the onset of significant phase decorrelation.

\begin{figure}[!t]
    \centering
    \includegraphics[width=8cm]{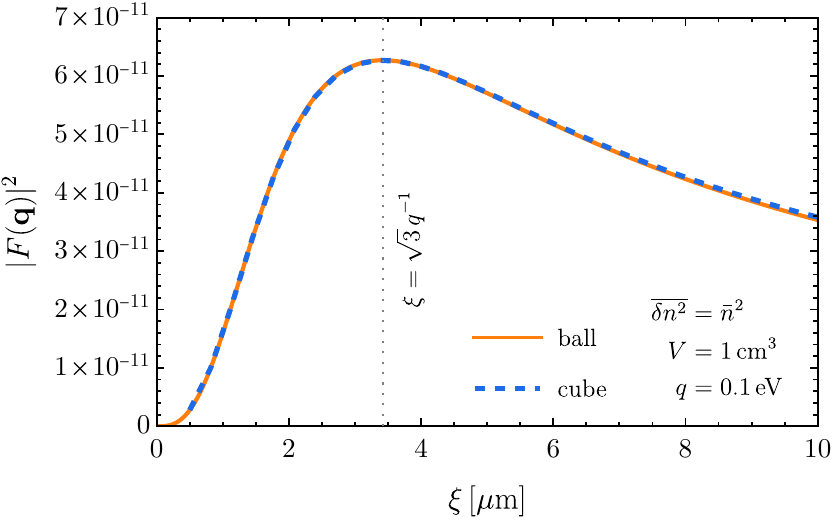}
    \caption{Form factors for different correlation lengths $\xi$ for spherical (solid) and cubic (dashed) targets with equal volume $V = 1\,\mathrm{cm}^3$ at a fixed momentum transfer $q = 0.1\,\mathrm{eV}$. The variance is set to $\overline{\delta n^2} = n_0^2$. For the cubic target, the momentum $\mathbf{q}$ is taken along one of its principal edges.}
    \label{fig:form_factor_geometry}
\end{figure}

For short-wavelength DM, the form factor $|F(\mathbf{q})|^{2}$ shows very little dependence on the overall target geometry. Physically, the DM ``sees'' the medium only over clusters of size $\xi \simeq q^{-1}$, and amplitudes from different clusters add incoherently. Consequently, the result is governed primarily by the underlying fluctuation statistics rather than by how clusters are assembled into macroscopic targets. This geometry insensitivity is confirmed numerically. As shown in Fig.~\ref{fig:form_factor_geometry}, at $q = 0.1\,\mathrm{eV}$ the form factors for spherical and cubic targets with the same volume $V = 1\,\mathrm{cm^{3}}$ are nearly indistinguishable.

\section{Acceleration from DM Scattering} 

Due to the rotation of the Solar System around the Galactic Center, there exists a relative velocity between the Solar System and the dark-matter (DM) halo, typically $|\mathbf{v}_\chi| \simeq 10^{-3}$ in the laboratory frame. With such a velocity, the energy transferred in sub-keV DM scattering lies far below the thresholds of conventional direct-detection experiments. However, light DM has a high number density, $n_\chi \equiv \rho_\chi / m_\chi$, where $\rho_\chi \simeq 0.4~\mathrm{GeV/cm}^3$ is the local DM density, and its scattering cross section with a macroscopic target can be coherently enhanced, yielding a sizable overall scattering rate. Continuous DM scattering can thus induce a measurable acceleration of the target, providing a unique observable for probing light DM~\cite{Day:2023mkb, Luo:2024ocg, Matsumoto:2025rcz}. Averaging over the DM phase space, the DM-induced acceleration of the target along the detection axis (taken to be the $z$-axis) is given by~\cite{Luo:2024ocg, Matsumoto:2025rcz}
\begin{equation}
    a_z = 
    \frac{n_\chi}{m_\mathrm{tot}} 
    \int d^3 \mathbf{v}_\chi\,d\Omega\,
    q_z\,|\mathbf{v}_\chi|\,
    f(\mathbf{v}_\chi)\,
    \left\langle \frac{d\sigma_{\rm tot}}{d\Omega} \right\rangle .
    \label{acc}
\end{equation}
Here, $m_\mathrm{tot}$ is the total target mass, $f(\mathbf{v}_\chi)$ is the DM velocity distribution in the laboratory frame~\cite{Baxter:2021pqo}, and $q_z$ is the $z$-component of the momentum transfer $\mathbf{q}$.

The observable scales as $a \sim \sigma_{\text{tot}} / m_{\text{tot}}$, where the total target mass is proportional to the number of atoms, $m_{\text{tot}} \propto N_\mathrm{tot}$. For sub-keV DM scattering off a random material, the total cross section behaves as $\sigma_{\text{tot}} \propto N_c N_\mathrm{tot}$, so the factor of $N_\mathrm{tot}$ cancels, yielding an additional enhancement by $N_c$ in the induced acceleration. Such accelerations can be sensitively probed in equivalence-principle tests, including the MICROSCOPE satellite mission~\cite{MICROSCOPE:2019jix, MICROSCOPE:2022doy} and torsion-balance experiments.

\section{Implementation in MICROSCOPE} 

The MICROSCOPE space mission was designed to test the weak equivalence principle (EP) by precisely monitoring the differential acceleration of two concentric cylindrical test masses in low-Earth orbit~\cite{MICROSCOPE:2019jix,MICROSCOPE:2022doy}. These test masses are made of two distinct materials, a Pt--Rh alloy and a Ti--6Al--4V alloy (90\% Ti, 6\% Al, 4\% V). The latter is a random two-phase alloy in which the $\alpha$ phase (hexagonal close-packed (HCP), Al-enriched) and the $\beta$ phase (body-centered cubic (BCC), V-enriched) coexist with different densities and compositions~\cite{Jaber:2022aaa}.

When sub-keV DM flows through the apparatus, it scatters off both test masses. The coherent enhancement of scattering on the uniform Pt--Rh alloy is suppressed compared to that on the Ti--6Al--4V alloy because of its random two-phase structure. As indicated by Eq.~\eqref{acc}, this difference in the total cross sections of the two test masses induces a measurable differential acceleration.

Since the exact microstructure of the Ti--6Al--4V alloy used in the MICROSCOPE experiment is not publicly available, we adopt as a benchmark the sample characterized in Ref.~\cite{Jaber:2022aaa}, whose lattice parameters and phase densities are summarized in Table~\ref{tab:Ti64PhaseSummary}. Together with the reported phase fractions, $f_\alpha = 74.39\%$ and $f_\beta = 25.61\%$~\cite{Jaber:2022aaa}, one can compute the average nucleon density and the variance of its density fluctuations as
\begin{equation}
    n_0 = 2.69 \times 10^{24}/\mathrm{cm^3} ,
    ~~
    \overline{\delta n^2} = \left( 7.29 \times 10^{22}/\mathrm{cm^3} \right)^2.
    \label{density}
\end{equation}
The detailed procedure for computing the density and its fluctuations from the material information is given in Appendix~\ref{appA}. In addition, the two-point correlation function of the Ti--6Al--4V alloy was measured experimentally in Ref.~\cite{Mahdavi:2019aaa}. The reported correlation function is well described by the simple exponential form in Eq.~\eqref{correlation_function_exponential}, with a characteristic correlation length $\xi$ of about $2\,\mu\mathrm{m}$.

\begin{table*}[!t]
    \centering
    \small
    \setlength{\tabcolsep}{5pt}
    \renewcommand{\arraystretch}{1.2}
    \begin{tabular}{ccccccc}
         \hline
            Phase    &  Structure & Lattice Parameter (\AA) & Weight Frac. (Ti/Al/V) & Volume Frac. &  Nucleon Number Density (cm$^{-3}$) \\
            \hline
            $\alpha$ & HCP & $a=2.9228$, $c=4.6692$ & 92.05/6.30/1.66  & 74.39\,\% & $2.65\times 10^{24}$ \\
            $\beta$  & BCC & $a=3.2001$             & 88.44/5.50/6.05  & 25.61\,\% & $2.81\times 10^{24}$ \\
            \hline
    \end{tabular}
    \caption{Summary of the microstructural properties of the $\alpha$ and $\beta$ phases in the Ti–6Al–4V alloy.}
    \label{tab:Ti64PhaseSummary}
\end{table*}

The MICROSCOPE test bodies are cylindrical shells whose size is much larger than the correlation length $\xi$, so the form factor is essentially insensitive to their global geometry. We therefore approximate them as solid spheres with the same mass in our analysis and employ Eq.~\eqref{form_factor_fluctuation} to evaluate the fluctuation-induced form factor.

\begin{figure}[!t]
    \centering
    \includegraphics[width=8cm]{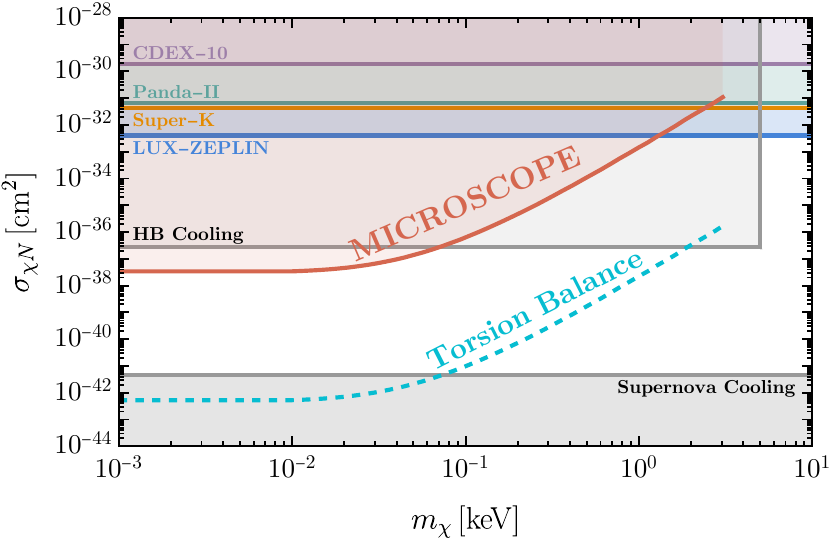}
    \caption{Constraints on the DM–nucleon scattering cross section from the MICROSCOPE mission (red) and a projected torsion-balance experiment (dashed cyan), compared with bounds from cosmic-ray–boosted DM and stellar-cooling.}
    \label{fig:limit}
\end{figure}

The DM-induced acceleration can be obtained by integrating \geqn{acc} under the assumption of an isotropic differential cross section, $d\sigma_{\chi N}/d\Omega = \sigma_{\chi N}/4\pi$. Due to the self-rotation of the MICROSCOPE satellite with frequency $f_s = 2.94 \times 10^{-3}\,\mathrm{Hz}$~\cite{MICROSCOPE:2019jix}, the DM-induced acceleration is modulated at a frequency $f_{\rm DM} = f_s$. This is close to the frequency of the equivalence-principle (EP)-violating signal induced by the Earth's gravity, $f_{\rm EP} = f_{o} + f_s = 3.11 \times 10^{-3}\,\mathrm{Hz}$, where $f_{o} = 1.68 \times 10^{-4}\,\mathrm{Hz}$ is the orbital frequency of the satellite~\cite{MICROSCOPE:2019jix}. Therefore, the constraints reported for EP violation, $\Delta a \lesssim 4.8 \times 10^{-14}\,\mathrm{m\,s^{-2}}$ at 95\% C.L.~\cite{MICROSCOPE:2022doy}, can be used as a benchmark for DM-induced acceleration. We find that the MICROSCOPE mission is sensitive to $\sigma_{\chi N}$ as small as $4 \times 10^{-38}\,\mathrm{cm^2}$ around $m_\chi \simeq 0.02\,\mathrm{keV}$, with the sensitivity gradually decreasing toward higher $m_\chi$, as shown in Fig.\,\ref{fig:limit}.

\section{Torsion Balance Experiment}
\label{sec:torsion}

Torsion-balance (TB) experiments provide the highest sensitivity to differential accelerations between test bodies, reaching sensitivities of order $10^{-15}\,\mathrm{m/s^2}$~\cite{Schlamminger:2007ht, Wagner:2012ui, Zhu:2018mrf}. However, test bodies in torsion-balance experiments are typically fabricated from compositionally homogeneous materials such as high-purity metals~\cite{Roll:1964rd, Braginskii:1971tn, Adelberger:1990xq, Su:1994gu, Schlamminger:2007ht, Wagner:2012ui} or quartz crystals~\cite{Zhu:2018mrf}. Such choices preclude taking advantage of the coherence enhancement we propose for random heterogeneous media and are therefore not optimal for efficiently probing light sub-keV DM scattering.

In future TB experiments, employing random heterogeneous materials with large density contrasts as test bodies can improve sensitivity to sub-keV DM scattering. One practical way to realize large density fluctuations with a controllable correlation length is to use commercially available porous tungsten. Porous tungsten, resembling a sponge with internal voids, is produced by pressing high-purity tungsten powder followed by high-temperature sintering. The correlation length can be tuned by varying the grain size of the precursor powder, while the number-density variance reaches its maximum when the pore volume fraction is around $50\%$, yielding
\begin{equation}
    \overline{\delta n^2} = n_0^2 = \left( 5.80 \times 10^{24} \, \mathrm{cm^{-3}} \right)^2.
\end{equation}
This variance is about four orders of magnitude larger than that of the Ti-6Al-4V alloy shown in Eq.~\eqref{density}.

When future TB experiments with test bodies of correlation length $\xi = 2 \, \mathrm{\mu m}$, as in MICROSCOPE, achieve an acceleration sensitivity of $\Delta a \sim 10^{-15} \, \mathrm{m\,s^{-2}}$~\cite{Schlamminger:2007ht, Wagner:2012ui, Zhu:2018mrf}, which is about one order of magnitude stronger than MICROSCOPE, the sensitivity to the DM--nucleon cross section could improve by around five orders of magnitude. This would correspond to a projected limit of
\begin{equation}
    \sigma_{\chi N}
    \lesssim
    5 \times 10^{-43} \, \mathrm{cm^2} .
    \label{sigma_TB1}
\end{equation}
This projection is shown as the dashed curve in Fig.~\ref{fig:limit}.

\section{Other Constraints} 

Although sub-keV DM carries too little kinetic energy to be detected in conventional direct-detection experiments, it can be boosted by cosmic rays to relativistic speeds, enabling elastic scattering off liquid-xenon targets and depositing recoil energies above threshold. Such cosmic-ray–boosted DM is constrained by PandaX-II~\cite{PandaX-II:2021kai}, CDEX-10~\cite{CDEX:2022fig}, LZ~\cite{LZ:2025iaw}, and Super-Kamiokande~\cite{Super-Kamiokande:2022ncz}. These limits are largely insensitive to the DM mass for two reasons: at smaller mass the DM number density is higher, while the energy transferred in each scattering is correspondingly smaller~\cite{Xia:2021vbz}.

In addition to direct detection, light DM is also constrained by astrophysical observations. The basic idea is that DM can be produced in stellar interiors through nucleon bremsstrahlung~\cite{Raffelt:1996wa}, Compton scattering~\cite{Raffelt:1996wa}, or electron--ion bremsstrahlung~\cite{Hardy:2016kme}, thereby carrying away energy and modifying stellar cooling. The most stringent bounds arise from supernova cooling. However, in supernova cores the nucleon density is so high that DM with sufficiently strong interactions becomes trapped and cannot escape, causing the supernova limits to saturate at large cross sections. On the other hand, strong interactions are constrained by horizontal branch (HB) stars, where DM emission mainly originates from electron--ion bremsstrahlung~\cite{Hardy:2016kme}, as shown by the light gray region in \gfig{fig:limit}. The evaluation of these stellar-cooling constraints is detailed in Appendix~\ref{appB}. Our mechanism is capable of probing the currently unconstrained region between the supernova limits and the HB-star constraints.

\section{Conclusion} 

The coherence effect can substantially enhance the scattering cross section of light DM on macroscopic targets. In this Letter, we demonstrate that random materials with small-scale density fluctuations exhibit an enhanced coherent response to sub-keV DM scattering, thereby amplifying the accelerations induced in the target by the continuous DM flux. We develop a general theoretical framework based on density–density correlation functions to compute the scattering form factor of random media. Within this framework, the MICROSCOPE experiment sets stringent constraints on sub-keV DM that surpass those from cosmic-ray–boosted DM and HB-star cooling, reaching $\sigma_{\chi N} \simeq 10^{-38}\,\mathrm{cm}^2$ for $m_\chi \simeq 10\,\mathrm{eV}$. Future torsion-balance experiments employing optimized random materials could further improve the sensitivity, allowing this new mechanism to probe regions of parameter space that were previously inaccessible.

\section*{Acknowledgements}

\noindent
S.~M. was supported by the Grant-in-Aid for Scientific Research from the Ministry of Education, Culture, Sports, Science and Technology, Japan (MEXT), under Grant Nos. 24H00244 and 24H02244.
J.~S. is supported by the Japan Society for the Promotion of Science (JSPS) as a part of the JSPS Postdoctoral Program (Standard) with grant number: P25018.
S.~M. and J.~S. are also supported by the World Premier International Research Center Initiative (WPI), MEXT, Japan (Kavli IPMU).
C.-Y. Xing is supported by the Fundamental Research Funds for the Central Universities (No.~24CX06048A). 

\begin{appendix}

\section{Density Variance of Ti-6Al-4V Alloy}
\label{appA}

The microstructure of the Ti-6Al-4V alloy under various heat-treatment conditions has been investigated in Ref.~\cite{Jaber:2022aaa}. As a representative example throughout this work, we consider the benchmark sample ``HT850WC+AG'', in which the alloy is heat-treated at $850^\circ$C for 2~hours, followed by water quenching and subsequent aging (reheating) at $500^\circ$C for 3~hours. The weight fractions of (Ti, Al, V) in the $\alpha$ phase are measured to be (92.05\%, 6.30\%, 1.66\%), while those in the $\beta$ phase are (88.44\%, 5.50\%, 6.05\%). In addition, the $\alpha$ phase has a hexagonal close-packed (HCP) structure with lattice parameters $a_\alpha = 2.9228\,\text{\AA}$ and $c_\alpha = 4.6692\,\text{\AA}$, whereas the $\beta$ phase has a body-centered cubic (BCC) structure with lattice parameter $a_\beta = 3.2001\,\text{\AA}$. The volumes of their primitive cells can be calculated as
\begin{subequations}
    \begin{align}
        V_{\mathrm{cell}, \alpha} &= \frac{\sqrt{3}}{2} a_\alpha^2 c_\alpha = 34.54 \, \text{\AA}^3, \\
        V_{\mathrm{cell}, \beta} &= a_\beta^3 = 32.77 \, \text{\AA}^3.
    \end{align}
    \label{volumes}
\end{subequations}

Using the atomic weights of $(m_\mathrm{Ti}, m_\mathrm{Al}, m_\mathrm{V}) = (47.867, 26.982, 50.942)$ in atomic mass units, one can calculate the densities of the two phases as follows:
\begin{equation}
    \rho
    = \frac{N_\mathrm{cell}}{V_\mathrm{cell}}
    ( f_{N, \mathrm{Ti}}\,m_\mathrm{Ti}
    + f_{N, \mathrm{Al}}\,m_\mathrm{Al}
    + f_{N, \mathrm{V}}\,m_\mathrm{V} ).
    \label{rho_by_nfrac}
\end{equation}
Here, $N_\mathrm{cell}$ denotes the number of atoms in the primitive unit cell. For both the two phases $\alpha$ and $\beta$, these are
\begin{equation}
    N_{\mathrm{cell}, \alpha} = N_{\mathrm{cell}, \beta} = 2.
\end{equation}
The volume of the primitive unit cell, $V_\mathrm{cell}$, is given in Eq.~\eqref{volumes}. Here, $f_{N,i}$ denotes the number fraction of element $i$ in the phase. It can be obtained from the weight fractions $f_{w,i}$ in each phase measured in Ref.~\cite{Jaber:2022aaa}. The mass of species $i$ in the material is $M_i = f_{w,i} M$, and the corresponding number of atoms is $N_i = f_{w,i} M / m_i$. The number fraction $f_{N,i}$ is defined as the ratio of $N_i$ to the total number of atoms, so that the total mass $M$ simply cancels. Consequently, the resulting relation between the number fractions $f_{N,i}$ and the weight fractions $f_{w,i}$ is
\begin{equation}
    f_{N, i} 
    = 
    \frac{f_{w, i} / m_i}{f_{w, \mathrm{Ti}} / m_\mathrm{Ti} + f_{w, \mathrm{Al}} / m_\mathrm{Al} + f_{w, \mathrm{V}} / m_\mathrm{V}}.
    \label{numberfrac}
\end{equation}
Substituting Eq.~\eqref{numberfrac} into Eq.~\eqref{rho_by_nfrac}, the density becomes
\begin{equation}
    \rho = \frac{N_\mathrm{cell}}{V_\mathrm{cell}}
    \frac{1}{f_{w, \mathrm{Ti}} / m_\mathrm{Ti} + f_{w, \mathrm{Al}} / m_\mathrm{Al} + f_{w, \mathrm{V}} / m_\mathrm{V}}.
\end{equation}
Using the measured parameters listed above, the densities of the $\alpha$ and $\beta$ phases are explicitly obtained as
\begin{subequations}
    \begin{align}
        \rho_\alpha &= 4.39 \, \mathrm{g/cm^3}, \\
        \rho_\beta &= 4.66 \, \mathrm{g/cm^3}.
    \end{align}
\end{subequations}
Because the primitive cell of the $\alpha$ phase has a larger volume and contains a higher proportion of aluminum atoms, the density is lower than that of the $\beta$ phase.

The volume fractions of the $\alpha$ and $\beta$ phases in the alloy are measured to be $74.39\%$ and $25.61\%$, respectively~\cite{Jaber:2022aaa}. The resulting average density of the alloy is given by
\begin{equation}
    \bar{\rho}
    =
    0.7439\,\rho_\alpha + 0.2561\,\rho_\beta
    = 4.46 \,\mathrm{g/cm^3} ,
\end{equation}
which is consistent with the density of the Ti-6Al-4V alloy, $4.42 \,\mathrm{g/cm^3}$, used in the MICROSCOPE mission~\cite{MICROSCOPE:2019jix}. Correspondingly, the nucleon number densities in the $\alpha$ and $\beta$ phases are then given approximately by
\begin{subequations}
    \begin{align}
        n_\alpha &= 2.65 \cdot 10^{24} \, \mathrm{cm^{-3}}, \\
        n_\beta &= 2.81 \cdot 10^{24} \, \mathrm{cm^{-3}}.
    \end{align}
\end{subequations}
The average nucleon number density in the alloy is,
\begin{equation}
    \bar{n}
    =
    0.7439 \, n_\alpha + 0.2561 \, n_\beta
    = 2.688 \cdot 10^{24} \, \mathrm{cm^{-3}},
\end{equation}
while the variance of the number density is
\begin{equation}
    \begin{split}
        \overline{\delta n^2}
        &= 0.7439 (n_\alpha - \bar{n})^2 + 0.2561 (n_\beta - \bar{n})^2 \\
        &= \left( 7.29 \cdot 10^{22} \, \mathrm{cm^{-3}} \right)^2.
    \end{split}
    \label{variance_Ti64}
\end{equation}
All relevant parameters of the two phases are summarized in \gtab{tab:Ti64PhaseSummary} in the main text for easy reference.

\section{Astrophysical Constraints}
\label{appB}

The constraints from stellar cooling relevant to this work arise from supernovae and horizontal-branch stars.

\vspace{0.2cm}
{\bf Supernova} -- Light DM that couples to nucleons can be produced in the core of supernovae via nucleon bremsstrahlung, thereby affecting the cooling process. However, if the interaction strength is sufficiently large, the produced DM particles become trapped within the core and can no longer efficiently transport energy away. In this trapping regime, the DM particles can be treated as blackbody radiation emitted from a DM-sphere inside the core. Inside the DM-sphere, the DM remains in thermal equilibrium with the surrounding medium, while outside it the DM free-streams out of the core. Then, the corresponding DM luminosity is given by~\cite{Raffelt:1996wa, Carenza:2019pxu},
\begin{equation}
    L_\chi 
    =
    \frac{g_\chi \pi}{120} 
    4\pi r_\mathrm{sp}^2 T_\mathrm{sp}^4 ,
\end{equation}
where $g_\chi$ is the number of degrees of freedom of the DM particle, and $r_\mathrm{sp}$ and $T_\mathrm{sp}$ denote the characteristic radius and temperature of the DM-sphere, respectively.

The radius of the DM-sphere is set by requiring that a DM particle undergo approximately one scattering while traveling from the DM-sphere to the core edge,
\begin{equation}
    \int_{r_\mathrm{sp}}^{\infty} \frac{dr}{\lambda_\chi} \simeq 1 ,
    \label{criterion}
\end{equation}
where the DM mean free path in the core is given by
\begin{equation}
    \lambda_\chi
    \equiv
    \frac{m_N}{\rho\,\sigma_{\chi N}} ,
\end{equation}
where $m_N$ is the nucleon mass, $\rho$ is the mass density of nucleons in the core, and $\sigma_{\chi N}$ is the DM scattering cross section with nucleons.
The density profile of the core can be approximated by a power-law function as~\cite{Turner:1987by, Raffelt:1996wa}
\begin{equation}
    \rho(r) = \rho_R \left( \frac{R}{r} \right)^\gamma ,
\end{equation}
where $\rho_R \simeq 10^{14} \, \mathrm{g/cm^3}$ is the density at $R \simeq 10 \, \mathrm{km}$, and the power-law index $\gamma$ typically lies in the range $3$–$7$. 
For a constant cross section $\sigma_{\chi N}$, integrating Eq.~\eqref{criterion} gives
\begin{equation}
    \sigma_{\chi N} \frac{\rho_R R^\gamma}{m_N}  \frac{r_\mathrm{sp}^{1-\gamma}}{\gamma-1}
    \simeq 1.
\end{equation}
The maximal scattering cross section for DM to escape from the DM-sphere can be analytically obtained as
\begin{equation}
    \sigma_{\chi N}
    \simeq
    \frac{(\gamma-1) m_N r_\mathrm{sp}^{\gamma-1} }{\rho_R R^\gamma}.
\end{equation}

The luminosity of new particles emitted from the supernova is constrained to be $L_\chi \lesssim 2 \times 10^{52} \, \mathrm{erg/s}$~\cite{Carenza:2019pxu}. Using this constraint, together with the local radial temperature profile of the protoneutron star reported in Ref.~\cite{Carenza:2019pxu}, we determine, in our benchmark scenario, the corresponding DM-sphere radius and temperature to be $r_\mathrm{sp} \simeq 19 \, \mathrm{km}$ and $T_\mathrm{sp} \simeq 7.5 \, \mathrm{MeV}$, respectively. This, in turn, yields a robust upper limit on the cross section,
\begin{equation}
    \sigma_{\chi N}
    \simeq 
    4.7 \times 10^{-42} \, \mathrm{cm^2},
\end{equation}
taking $\gamma = 7$ as a benchmark. Dark matter with a cross section larger than this value is trapped in the core and is therefore not constrained by supernova cooling.

\vspace{0.2cm}
{\bf Horizontal-Branch (HB) Stars} -- 
Dedicated studies of HB-star cooling constraints on quadratically coupled light dark matter are not available. Nevertheless, for linearly coupled scalar particles $\phi$ with interaction $g \phi \bar N N$, the cooling effects have been extensively studied in detail~\cite{Hardy:2016kme}; the production of $\phi$ is dominated by electron–ion bremsstrahlung in HB stars, and the upper bound on the coupling is $g^2/4\pi \simeq 5 \times 10^{-23}$ for $m_\phi \lesssim 10\,\mathrm{keV}$~\cite{Hardy:2016kme}. We translate this bound into the corresponding constraint on the quadratically coupled light dark matter $\chi$ considered here by comparing their production rates in HB stars.

The thermal energy in HB stars is of order $10\,\mathrm{keV}$, which is much larger than the DM mass but much smaller than both electron and nucleon masses. Consequently, the emitted DM particles can be treated as relativistic, whereas the electrons and nucleons participating in the scattering process are nonrelativistic. In this regime, the matrix element for single-$\phi$ bremsstrahlung emission can be approximated as $\mathcal{M}_{\phi} \simeq (i g/\omega_k)\,\mathcal{M}_0$, where $\mathcal{M}_0$ is the amplitude for the underlying electron or nucleon interaction without final-state emission, and $\omega_k$ is the energy of the emitted $\phi$ with four-momentum $k \simeq (\omega_k,{\bf k})$. The prefactor $i g/\omega_k$ arises from the fermion propagator. This approximation is valid as long as $\omega_k \ll m_N$.
When the scalar $\phi$ is soft, the nucleon propagator is almost on-shell and the cross section of $\phi$-emission can be factorized as
\begin{equation}
    d\sigma_{\phi} 
    \approx 
    d\sigma_{0} 
    \times 
    \frac{g^2}{\omega_k^2}
    \frac{d^3 \mathbf{k}}{(2\pi)^3 2 \omega_k},
\label{cs_phi}
\end{equation}
where $d\sigma_{0}$ is the cross section of electron-ion interaction without the final state emission. The energy-loss rate in HB stars is proportional to the thermal average $\langle \omega_k\,\sigma\,v_{\rm rel}\rangle$. For single-scalar emission, it is given by
\begin{equation}
    \begin{aligned}
        \langle \omega_k \sigma_{\phi} v_\mathrm{rel} \rangle
        &\simeq
        \sigma_{0} v_\mathrm{rel} \times
        \int \frac{g^2}{\omega_k^2} \omega_k \frac{d^3 \mathbf{k}}{(2\pi)^3 2\omega_k}  \\
        &=
        \sigma_{0} v_\mathrm{rel} \times
        \frac{g^2}{2\pi^2} E_{\max},
    \end{aligned}
\end{equation}
where $E_{\max}$ denotes the maximum energy of the emitted $\phi$, which is of order $10\,\mathrm{keV}$ in typical HB stars.

Now, we compute the energy-emission rate of quadratically coupled DM $\chi$ with interaction $\chi^2 \bar N N / f_N$ in HB stars, and derive the corresponding constraint by relating it to that of the scalar field $\phi$; under similar assumptions as before, the cross section can be approximated as,
\begin{equation}
    d\sigma_{2 \chi} 
    \approx 
    d\sigma_{0}  
    \frac{1}{f_N^2 (\omega_{k_1} + \omega_{k_2})^2}
    \frac{d^3 \mathbf{k}_1}{(2\pi)^3 2 \omega_{k_1}}  
    \frac{d^3 \mathbf{k}_2}{(2\pi)^3 2 \omega_{k_2}}  .
\end{equation}
Compared with Eq.~\eqref{cs_phi}, the only difference is that the final state now contains two DM particles with momenta $k_1$ and $k_2$, so the momentum flowing through the fermion propagator becomes $(k_1+k_2)$. The corresponding energy-emission rate in HB stars is then given as follows:
\begin{equation}
    \langle(
        \omega_{k_1} + \omega_{k_2}) \sigma_{2 \phi} v_\mathrm{rel}
    \rangle
    \simeq 
    \sigma_{0} v_\mathrm{rel} \times \mathcal{I},
\end{equation}
where the integration represented by $\mathcal{I}$ is defined as,
\begin{align}
    \mathcal{I}
    &\equiv
    \int \frac{1}{f_N^2(\omega_{k_1}+\omega_{k_2})^2} (\omega_{k_1}+\omega_{k_2})
    \frac{d^3\mathbf{k}_1}{(2\pi)^3\,2\omega_{k_1}}
    \frac{d^3\mathbf{k}_2}{(2\pi)^3\,2\omega_{k_2}}
    \nonumber \\
    &=
    \frac{1}{16\pi^4 f_N^2}
    \int \int \frac{\omega_1\omega_2}{\omega_1+\omega_2}\,d\omega_1 d\omega_2
    =
    \frac{E_{\max}^3}{288\pi^4 f_N^2}.
\end{align}
The maximum energy released is still $E_{\max}\sim T\sim 10\,\mathrm{keV}$ in HB stars, set by the temperature. The ratio of the energy-emission rates in the $\phi$ and $\chi$ scenarios is
\begin{equation}
    \frac{\Gamma_{\phi}}{\Gamma_{2 \chi}}
    \sim \frac{g^2 E_{\max} / (2\pi^2)}{E_{\max}^3/(288\pi^4 f_N^2)}
    = \frac{144\pi^2 g^2 f_N^2}{E_{\max}^2}.
\end{equation}
The HB constraint on the $\phi$ coupling is $g^2/(4\pi) < 5 \times 10^{-23}$; this can be translated into a constraint on $f_N$ as
\begin{equation}
    \frac{1}{f_N^2}
    < 4\pi \frac{144\pi^2}{E_{\max}^2} 5 \times 10^{-23}
    \approx \frac{1}{(10.6 \, \mathrm{TeV})^2} 
    \left(\frac{10 \, \mathrm{keV}}{E_{\max}}\right)^2.
\end{equation}
Therefore, the corresponding bound on the scattering cross section between $\chi$ and a nucleon $N$ is given by
\begin{equation}
    \sigma_{\chi N} 
    = \frac{1}{4\pi f_N^2} 
    \simeq 2.77 \times 10^{-37} \, \mathrm{cm}^2
    \left(\frac{10 \, \mathrm{keV}}{E_{\max}}\right)^2.
\end{equation}

\end{appendix}

\bibliographystyle{utphys}
\bibliography{ref}

@article{Baxter:2021pqo,
    author = "Baxter, D. and others",
    title = "{Recommended conventions for reporting results from direct dark matter searches}",
    eprint = "2105.00599",
    archivePrefix = "arXiv",
    primaryClass = "hep-ex",
    doi = "10.1140/epjc/s10052-021-09655-y",
    journal = "Eur. Phys. J. C",
    volume = "81",
    number = "10",
    pages = "907",
    year = "2021"
}

@article{Jiao:2007aaa,
    author = "Jiao, Y. and Stillinger, F. H. and Torquato, S.",
    title = "{Modeling heterogeneous materials via two-point correlation functions: Basic principles}",
    eprint = "0705.2434",
    archivePrefix = "arXiv",
    doi = {10.1103/PhysRevE.76.031110},
    journal = "Phys. Rev. E",
    number = "3",
    volume = "76",
    pages = "031110",
    year = "2007"
}

@article{Jaber:2022aaa,
    author = "Jaber, H. and Kónya, J. and Kulcsár, K. and Kovács, T.",
    title = "{Effects of Annealing and Solution Treatments on the Microstructure and Mechanical Properties of Ti6Al4V Manufactured by Selective Laser Melting}",
    doi = {10.3390/ma15051978},
    journal = "Materials (Basel, Switzerland)",
    volume = "15",
    number = "5",
    pages = "1978",
    year = "2022"
}

@article{MICROSCOPE:2019jix,
    author = "Touboul, Pierre and others",
    collaboration = "MICROSCOPE",
    title = "{Space test of the Equivalence Principle: first results of the MICROSCOPE mission}",
    eprint = "1909.10598",
    archivePrefix = "arXiv",
    primaryClass = "gr-qc",
    doi = "10.1088/1361-6382/ab4707",
    journal = "Class. Quant. Grav.",
    volume = "36",
    number = "22",
    pages = "225006",
    year = "2019"
}

@article{Mahdavi:2019aaa,
    author = {Mostafa Mahdavi and Eric Hoar and Daniel E. Sievers and Yan Chong and Nobuhiro Tsuji and Steven Liang and Hamid Garmestani},
    title = {Statistical representation of the microstructure and strength for a two-phase Ti-6Al-4V},
    journal = {Materials Science and Engineering: A},
    volume = {759},
    pages = {313-319},
    year = {2019},
    issn = {0921-5093},
    doi = {https://doi.org/10.1016/j.msea.2019.05.048}
}

@article{MICROSCOPE:2022doy,
    author = "Touboul, Pierre and others",
    collaboration = "MICROSCOPE",
    title = "{MICROSCOPE Mission: Final Results of the Test of the Equivalence Principle}",
    eprint = "2209.15487",
    archivePrefix = "arXiv",
    primaryClass = "gr-qc",
    doi = "10.1103/PhysRevLett.129.121102",
    journal = "Phys. Rev. Lett.",
    volume = "129",
    number = "12",
    pages = "121102",
    year = "2022"
}

@book{Raffelt:1996wa,
    author = "Raffelt, G. G.",
    title = "{Stars as laboratories for fundamental physics}: {The astrophysics of neutrinos, axions, and other weakly interacting particles}",
    isbn = "978-0-226-70272-8",
    month = "5",
    year = "1996",
    publisher = "University of Chicago Press"
}

@article{Carenza:2019pxu,
    author = "Carenza, Pierluca and Fischer, Tobias and Giannotti, Maurizio and Guo, Gang and Mart{\'\i}nez-Pinedo, Gabriel and Mirizzi, Alessandro",
    title = "{Improved axion emissivity from a supernova via nucleon-nucleon bremsstrahlung}",
    eprint = "1906.11844",
    archivePrefix = "arXiv",
    primaryClass = "hep-ph",
    doi = "10.1088/1475-7516/2019/10/016",
    journal = "JCAP",
    volume = "10",
    number = "10",
    pages = "016",
    year = "2019",
    note = "[Erratum: JCAP 05, E01 (2020)]"
}

@article{Turner:1987by,
    author = "Turner, Michael S.",
    title = "{Axions from SN 1987a}",
    reportNumber = "FERMILAB-PUB-87-202-A",
    doi = "10.1103/PhysRevLett.60.1797",
    journal = "Phys. Rev. Lett.",
    volume = "60",
    pages = "1797",
    year = "1988"
}

@article{LZ:2025iaw,
    author = "Aalbers, J. and others",
    collaboration = "LZ",
    title = "{New Constraints on Cosmic Ray-Boosted Dark Matter from the LUX-ZEPLIN Experiment}",
    eprint = "2503.18158",
    archivePrefix = "arXiv",
    primaryClass = "hep-ex",
    reportNumber = "FERMILAB-PUB-25-0191-V",
    doi = "10.1103/nr92-jvt3",
    journal = "Phys. Rev. Lett.",
    volume = "134",
    number = "24",
    pages = "241801",
    year = "2025"
}

@article{Super-Kamiokande:2022ncz,
    author = "Abe, K. and others",
    collaboration = "Super-Kamiokande",
    title = "{Search for Cosmic-Ray Boosted Sub-GeV Dark Matter Using Recoil Protons at Super-Kamiokande}",
    eprint = "2209.14968",
    archivePrefix = "arXiv",
    primaryClass = "hep-ex",
    doi = "10.1103/PhysRevLett.130.031802",
    journal = "Phys. Rev. Lett.",
    volume = "130",
    number = "3",
    pages = "031802",
    year = "2023",
    note = "[Erratum: Phys.Rev.Lett. 131, 159903 (2023)]"
}

@article{PandaX-II:2021kai,
    author = "Cui, Xiangyi and others",
    collaboration = "PandaX-II",
    title = "{Search for Cosmic-Ray Boosted Sub-GeV Dark Matter at the PandaX-II Experiment}",
    eprint = "2112.08957",
    archivePrefix = "arXiv",
    primaryClass = "hep-ex",
    doi = "10.1103/PhysRevLett.128.171801",
    journal = "Phys. Rev. Lett.",
    volume = "128",
    number = "17",
    pages = "171801",
    year = "2022"
}

@article{CDEX:2022fig,
    author = "Xu, R. and others",
    collaboration = "CDEX",
    title = "{Constraints on sub-GeV dark matter boosted by cosmic rays from the CDEX-10 experiment at the China Jinping Underground Laboratory}",
    eprint = "2201.01704",
    archivePrefix = "arXiv",
    primaryClass = "hep-ex",
    doi = "10.1103/PhysRevD.106.052008",
    journal = "Phys. Rev. D",
    volume = "106",
    number = "5",
    pages = "052008",
    year = "2022"
}

@article{Wagner:2012ui,
    author = "Wagner, T. A. and Schlamminger, S. and Gundlach, J. H. and Adelberger, E. G.",
    title = "{Torsion-balance tests of the weak equivalence principle}",
    eprint = "1207.2442",
    archivePrefix = "arXiv",
    primaryClass = "gr-qc",
    doi = "10.1088/0264-9381/29/18/184002",
    journal = "Class. Quant. Grav.",
    volume = "29",
    pages = "184002",
    year = "2012"
}

@article{Zhu:2018mrf,
    author = "Zhu, Lin and Liu, Qi and Zhao, Hui-Hui and Gong, Qi-Long and Yang, Shan-Qing and Luo, Pengshun and Shao, Cheng-Gang and Wang, Qing-Lan and Tu, Liang-Cheng and Luo, Jun",
    title = "{Test of the Equivalence Principle with Chiral Masses Using a Rotating Torsion Pendulum}",
    doi = "10.1103/PhysRevLett.121.261101",
    journal = "Phys. Rev. Lett.",
    volume = "121",
    number = "26",
    pages = "261101",
    year = "2018"
}

@article{Hardy:2016kme,
    author = "Hardy, Edward and Lasenby, Robert",
    title = "{Stellar cooling bounds on new light particles: plasma mixing effects}",
    eprint = "1611.05852",
    archivePrefix = "arXiv",
    primaryClass = "hep-ph",
    doi = "10.1007/JHEP02(2017)033",
    journal = "JHEP",
    volume = "02",
    pages = "033",
    year = "2017"
}

@article{Planck:2018vyg,
    author = "Aghanim, N. and others",
    collaboration = "Planck",
    title = "{Planck 2018 results. VI. Cosmological parameters}",
    eprint = "1807.06209",
    archivePrefix = "arXiv",
    primaryClass = "astro-ph.CO",
    doi = "10.1051/0004-6361/201833910",
    journal = "Astron. Astrophys.",
    volume = "641",
    pages = "A6",
    year = "2020",
    note = "[Erratum: Astron.Astrophys. 652, C4 (2021)]"
}

@article{Debye_1957, 
    title={Scattering by an Inhomogeneous Solid. II. The Correlation Function and Its Application}, 
    volume={28}, 
    ISSN={1089-7550}, 
    url={http://dx.doi.org/10.1063/1.1722830}, 
    DOI={10.1063/1.1722830}, 
    number={6}, 
    journal={Journal of Applied Physics}, 
    publisher={AIP Publishing}, 
    author={Debye, P. and Anderson, H. R. and Brumberger, H.}, 
    year={1957}, 
    month=jun, 
    pages={679–683} 
}

@article{Debye_1949, 
    title={Scattering by an Inhomogeneous Solid}, 
    volume={20}, 
    ISSN={1089-7550}, 
    url={http://dx.doi.org/10.1063/1.1698419}, 
    DOI={10.1063/1.1698419}, 
    number={6}, 
    journal={Journal of Applied Physics}, 
    publisher={AIP Publishing}, 
    author={Debye, P. and Bueche, A. M.}, 
    year={1949}, 
    month=jun, 
    pages={518–525} 
}

@article{Schlamminger:2007ht,
    author = "Schlamminger, Stephan and Choi, K. -Y. and Wagner, T. A. and Gundlach, J. H. and Adelberger, E. G.",
    title = "{Test of the equivalence principle using a rotating torsion balance}",
    eprint = "0712.0607",
    archivePrefix = "arXiv",
    primaryClass = "gr-qc",
    doi = "10.1103/PhysRevLett.100.041101",
    journal = "Phys. Rev. Lett.",
    volume = "100",
    pages = "041101",
    year = "2008"
}

@article{Roll:1964rd,
    author = "Roll, P. G. and Krotkov, R. and Dicke, R. H.",
    title = "{The Equivalence of inertial and passive gravitational mass}",
    doi = "10.1016/0003-4916(64)90259-3",
    journal = "Annals Phys.",
    volume = "26",
    pages = "442--517",
    year = "1964"
}

@article{Braginskii:1971tn,
    author = "Braginskii, V. B. and Panov, V. I.",
    title = "{Verification of equivalence of inertial and gravitational masses}",
    journal = "Sov.Phys.JETP",
    volume = "34",
    pages = "463–476",
    year = "1972"
}

@article{Adelberger:1990xq,
    author = "Adelberger, E. G. and Stubbs, C. W. and Heckel, Blayne R. and Su, Y. and Swanson, H. E. and Smith, G. and Gundlach, J. H. and Rogers, W. F.",
    title = "{Testing the equivalence principle in the field of the earth: Particle physics at masses below 1-microEV?}",
    doi = "10.1103/PhysRevD.42.3267",
    journal = "Phys. Rev. D",
    volume = "42",
    pages = "3267--3292",
    year = "1990"
}

@article{Su:1994gu,
    author = "Su, Y. and Heckel, Blayne R. and Adelberger, E. G. and Gundlach, J. H. and Harris, M. and Smith, G. L. and Swanson, H. E.",
    title = "{New tests of the universality of free fall}",
    doi = "10.1103/PhysRevD.50.3614",
    journal = "Phys. Rev. D",
    volume = "50",
    pages = "3614--3636",
    year = "1994"
}

@article{Tremaine:1979we,
    author = "Tremaine, S. and Gunn, J. E.",
    editor = "Srednicki, M. A.",
    title = "{Dynamical Role of Light Neutral Leptons in Cosmology}",
    doi = "10.1103/PhysRevLett.42.407",
    journal = "Phys. Rev. Lett.",
    volume = "42",
    pages = "407--410",
    year = "1979"
}

@article{Domcke:2014kla,
    author = "Domcke, Valerie and Urbano, Alfredo",
    title = "{Dwarf spheroidal galaxies as degenerate gas of free fermions}",
    eprint = "1409.3167",
    archivePrefix = "arXiv",
    primaryClass = "hep-ph",
    reportNumber = "SISSA-48-2014-FISI",
    doi = "10.1088/1475-7516/2015/01/002",
    journal = "JCAP",
    volume = "01",
    pages = "002",
    year = "2015"
}

@article{Alvey:2020xsk,
    author = "Alvey, James and Sabti, Nashwan and Tiki, Victoria and Blas, Diego and Bondarenko, Kyrylo and Boyarsky, Alexey and Escudero, Miguel and Fairbairn, Malcolm and Orkney, Matthew and Read, Justin I.",
    title = "{New constraints on the mass of fermionic dark matter from dwarf spheroidal galaxies}",
    eprint = "2010.03572",
    archivePrefix = "arXiv",
    primaryClass = "hep-ph",
    reportNumber = "KCL-2020-58, TUM-HEP-1285/20",
    doi = "10.1093/mnras/staa3640",
    journal = "Mon. Not. Roy. Astron. Soc.",
    volume = "501",
    number = "1",
    pages = "1188--1201",
    year = "2021"
}

@article{Bertone:2016nfn,
    author = "Bertone, Gianfranco and Hooper, Dan",
    title = "{History of dark matter}",
    eprint = "1605.04909",
    archivePrefix = "arXiv",
    primaryClass = "astro-ph.CO",
    reportNumber = "FERMILAB-PUB-16-157-A",
    doi = "10.1103/RevModPhys.90.045002",
    journal = "Rev. Mod. Phys.",
    volume = "90",
    number = "4",
    pages = "045002",
    year = "2018"
}

@article{Bertone:2004pz,
    author = "Bertone, Gianfranco and Hooper, Dan and Silk, Joseph",
    title = "{Particle dark matter: Evidence, candidates and constraints}",
    eprint = "hep-ph/0404175",
    archivePrefix = "arXiv",
    reportNumber = "FERMILAB-PUB-04-047-A",
    doi = "10.1016/j.physrep.2004.08.031",
    journal = "Phys. Rept.",
    volume = "405",
    pages = "279--390",
    year = "2005"
}

@article{Schumann:2019eaa,
    author = "Schumann, Marc",
    title = "{Direct Detection of WIMP Dark Matter: Concepts and Status}",
    eprint = "1903.03026",
    archivePrefix = "arXiv",
    primaryClass = "astro-ph.CO",
    doi = "10.1088/1361-6471/ab2ea5",
    journal = "J. Phys. G",
    volume = "46",
    number = "10",
    pages = "103003",
    year = "2019"
}

@article{Billard:2021uyg,
    author = "Billard, Julien and others",
    title = "{Direct detection of dark matter{\textemdash}APPEC committee report*}",
    eprint = "2104.07634",
    archivePrefix = "arXiv",
    primaryClass = "hep-ex",
    doi = "10.1088/1361-6633/ac5754",
    journal = "Rept. Prog. Phys.",
    volume = "85",
    number = "5",
    pages = "056201",
    year = "2022"
}

@article{Goodman:1984dc,
    author = "Goodman, Mark W. and Witten, Edward",
    editor = "Srednicki, M. A.",
    title = "{Detectability of Certain Dark Matter Candidates}",
    reportNumber = "Print-85-0030 (PRINCETON)",
    doi = "10.1103/PhysRevD.31.3059",
    journal = "Phys. Rev. D",
    volume = "31",
    pages = "3059",
    year = "1985"
}

@article{Drukier:1986tm,
    author = "Drukier, A. K. and Freese, Katherine and Spergel, D. N.",
    title = "{Detecting Cold Dark Matter Candidates}",
    doi = "10.1103/PhysRevD.33.3495",
    journal = "Phys. Rev. D",
    volume = "33",
    pages = "3495--3508",
    year = "1986"
}

@article{Lewin:1995rx,
    author = "Lewin, J. D. and Smith, P. F.",
    title = "{Review of mathematics, numerical factors, and corrections for dark matter experiments based on elastic nuclear recoil}",
    reportNumber = "RAL-TR-95-024",
    doi = "10.1016/S0927-6505(96)00047-3",
    journal = "Astropart. Phys.",
    volume = "6",
    pages = "87--112",
    year = "1996"
}

@article{Guo:2021imc,
    author = "Guo, Ji-Heng and Sun, Yu-Xuan and Wang, Wenyu and Wu, Ke-Yun",
    title = "{Can sub-GeV dark matter coherently scatter on the electrons in the atom?}",
    eprint = "2112.11810",
    archivePrefix = "arXiv",
    primaryClass = "hep-ph",
    doi = "10.1088/1572-9494/ac9f0b",
    journal = "Commun. Theor. Phys.",
    volume = "75",
    number = "1",
    pages = "015201",
    year = "2023"
}

@article{Luo:2024ocg,
    author = "Luo, Pengshun and Matsumoto, Shigeki and Sheng, Jie and Xing, Chuan-Yang and Zhu, Lin and Zhuge, Zhi-Jie",
    title = "{Detecting meV-scale dark matter via coherent scattering with an asymmetric torsion balance}",
    eprint = "2409.09950",
    archivePrefix = "arXiv",
    primaryClass = "hep-ph",
    doi = "10.1103/nhcn-9p3j",
    journal = "Phys. Rev. D",
    volume = "112",
    number = "7",
    pages = "075023",
    year = "2025"
}

@article{Matsumoto:2025rcz,
    author = "Matsumoto, Shigeki and Sheng, Jie and Xing, Chuan-Yang and Zhu, Lin",
    title = "{Torsion Balance Experiments Enable Direct Detection of Sub-eV Dark Matter}",
    eprint = "2506.07763",
    archivePrefix = "arXiv",
    primaryClass = "hep-ph",
    month = "6",
    year = "2025"
}

@article{Xia:2021vbz,
    author = "Xia, Chen and Xu, Yan-Hao and Zhou, Yu-Feng",
    title = "{Production and attenuation of cosmic-ray boosted dark matter}",
    eprint = "2111.05559",
    archivePrefix = "arXiv",
    primaryClass = "hep-ph",
    doi = "10.1088/1475-7516/2022/02/028",
    journal = "JCAP",
    volume = "02",
    number = "02",
    pages = "028",
    year = "2022"
}

@book{Bauer:2017qwy,
    author = "Bauer, Martin and Plehn, Tilman",
    title = "{Yet Another Introduction to Dark Matter}: {The Particle Physics Approach}",
    eprint = "1705.01987",
    archivePrefix = "arXiv",
    primaryClass = "hep-ph",
    doi = "10.1007/978-3-030-16234-4",
    publisher = "Springer",
    series = "Lecture Notes in Physics",
    volume = "959",
    year = "2019"
}

@article{PandaX:2024qfu,
    author = "Bo, Zihao and others",
    collaboration = "PandaX",
    title = "{Dark Matter Search Results from 1.54{\,}{\,}Tonne{\textperiodcentered}Year Exposure of PandaX-4T}",
    eprint = "2408.00664",
    archivePrefix = "arXiv",
    primaryClass = "hep-ex",
    doi = "10.1103/PhysRevLett.134.011805",
    journal = "Phys. Rev. Lett.",
    volume = "134",
    number = "1",
    pages = "011805",
    year = "2025"
}

@article{CDEX:2022kcd,
    author = "Zhang, Z. Y. and others",
    collaboration = "CDEX",
    title = "{Constraints on Sub-GeV Dark Matter{\textendash}Electron Scattering from the CDEX-10 Experiment}",
    eprint = "2206.04128",
    archivePrefix = "arXiv",
    primaryClass = "hep-ex",
    doi = "10.1103/PhysRevLett.129.221301",
    journal = "Phys. Rev. Lett.",
    volume = "129",
    number = "22",
    pages = "221301",
    year = "2022"
}

@article{SENSEI:2023zdf,
    author = "Adari, Prakruth and others",
    collaboration = "SENSEI",
    title = "{First Direct-Detection Results on Sub-GeV Dark Matter Using the SENSEI Detector at SNOLAB}",
    eprint = "2312.13342",
    archivePrefix = "arXiv",
    primaryClass = "astro-ph.CO",
    reportNumber = "YITP-SB-2023-30, FERMILAB-PUB-23-0824-CSAID-PPD",
    doi = "10.1103/PhysRevLett.134.011804",
    journal = "Phys. Rev. Lett.",
    volume = "134",
    number = "1",
    pages = "011804",
    year = "2025"
}

@article{Fukuda:2018omk,
    author = "Fukuda, Hajime and Matsumoto, Shigeki and Yanagida, Tsutomu T.",
    title = "{Direct Detection of Ultralight Dark Matter via Astronomical Ephemeris}",
    eprint = "1801.02807",
    archivePrefix = "arXiv",
    primaryClass = "hep-ph",
    reportNumber = "IPMU18-0003",
    doi = "10.1016/j.physletb.2018.12.038",
    journal = "Phys. Lett. B",
    volume = "789",
    pages = "220--227",
    year = "2019"
}

@article{Akhmedov:2018wlf,
    author = "Akhmedov, Evgeny and Arcadi, Giorgio and Lindner, Manfred and Vogl, Stefan",
    title = "{Coherent scattering and macroscopic coherence: Implications for neutrino, dark matter and axion detection}",
    eprint = "1806.10962",
    archivePrefix = "arXiv",
    primaryClass = "hep-ph",
    doi = "10.1007/JHEP10(2018)045",
    journal = "JHEP",
    volume = "10",
    pages = "045",
    year = "2018"
}

@article{Shergold:2021evs,
    author = "Shergold, Jack D.",
    title = "{Updated detection prospects for relic neutrinos using coherent scattering}",
    eprint = "2109.07482",
    archivePrefix = "arXiv",
    primaryClass = "hep-ph",
    reportNumber = "IPPP/21/28",
    doi = "10.1088/1475-7516/2021/11/052",
    journal = "JCAP",
    volume = "11",
    number = "11",
    pages = "052",
    year = "2021"
}

@article{Afek:2021vjy,
    author = "Afek, Gadi and Carney, Daniel and Moore, David C.",
    title = "{Coherent Scattering of Low Mass Dark Matter from Optically Trapped Sensors}",
    eprint = "2111.03597",
    archivePrefix = "arXiv",
    primaryClass = "physics.ins-det",
    doi = "10.1103/PhysRevLett.128.101301",
    journal = "Phys. Rev. Lett.",
    volume = "128",
    number = "10",
    pages = "101301",
    year = "2022"
}

@article{Fukuda:2021drn,
    author = "Fukuda, Hajime and Shirai, Satoshi",
    title = "{Detection of QCD axion dark matter by coherent scattering}",
    eprint = "2112.13536",
    archivePrefix = "arXiv",
    primaryClass = "hep-ph",
    reportNumber = "IPMU21-0089",
    doi = "10.1103/PhysRevD.105.095030",
    journal = "Phys. Rev. D",
    volume = "105",
    number = "9",
    pages = "095030",
    year = "2022"
}

@article{DarkSide-20k:2024yfq,
    author = "Acerbi, F. and others",
    collaboration = "DarkSide-20k",
    title = "{DarkSide-20k sensitivity to light dark matter particles}",
    eprint = "2407.05813",
    archivePrefix = "arXiv",
    primaryClass = "hep-ex",
    reportNumber = "FERMILAB-PUB-24-0483-V",
    doi = "10.1038/s42005-024-01896-z",
    journal = "Commun. Phys.",
    volume = "7",
    number = "1",
    pages = "422",
    year = "2024"
}

@article{SuperCDMS:2024yiv,
    author = "Albakry, M. F. and others",
    collaboration = "SuperCDMS",
    title = "{Light dark matter constraints from SuperCDMS HVeV detectors operated underground with an anticoincidence event selection}",
    eprint = "2407.08085",
    archivePrefix = "arXiv",
    primaryClass = "hep-ex",
    reportNumber = "FERMILAB-PUB-24-0376-PPD",
    doi = "10.1103/PhysRevD.111.012006",
    journal = "Phys. Rev. D",
    volume = "111",
    number = "1",
    pages = "012006",
    year = "2025"
}

@article{XENON:2025vwd,
    author = "Aprile, E. and others",
    collaboration = "XENON",
    title = "{WIMP Dark Matter Search Using a 3.1 Tonne-Year Exposure of the XENONnT Experiment}",
    eprint = "2502.18005",
    archivePrefix = "arXiv",
    primaryClass = "hep-ex",
    doi = "10.1103/msw4-t342",
    journal = "Phys. Rev. Lett.",
    volume = "135",
    number = "22",
    pages = "221003",
    year = "2025"
}

@article{DAMIC-M:2025luv,
    author = "Aggarwal, K. and others",
    collaboration = "DAMIC-M",
    title = "{Probing Benchmark Models of Hidden-Sector Dark Matter with DAMIC-M}",
    eprint = "2503.14617",
    archivePrefix = "arXiv",
    primaryClass = "hep-ex",
    doi = "10.1103/2tcc-bqck",
    journal = "Phys. Rev. Lett.",
    volume = "135",
    number = "7",
    pages = "071002",
    year = "2025"
}

@article{LZ:2024zvo,
    author = "Aalbers, J. and others",
    collaboration = "LZ",
    title = "{Dark Matter Search Results from 4.2{\,}{\,}Tonne-Years of Exposure of the LUX-ZEPLIN (LZ) Experiment}",
    eprint = "2410.17036",
    archivePrefix = "arXiv",
    primaryClass = "hep-ex",
    reportNumber = "FERMILAB-PUB-24-0796-V",
    doi = "10.1103/4dyc-z8zf",
    journal = "Phys. Rev. Lett.",
    volume = "135",
    number = "1",
    pages = "011802",
    year = "2025"
}

@article{Day:2023mkb,
    author = "Day, Hannah and Liu, Da and Luty, Markus A. and Zhao, Yue",
    title = "{Blowing in the dark matter wind}",
    eprint = "2312.13345",
    archivePrefix = "arXiv",
    primaryClass = "hep-ph",
    doi = "10.1007/JHEP07(2024)136",
    journal = "JHEP",
    volume = "07",
    pages = "136",
    year = "2024"
}

@article{Acevedo:2025rqu,
    author = "Acevedo, Javier F. and Reilly, Aidan J. and Santos-Olmsted, Lillian",
    title = "{Dark Drag Around Sagittarius A*}",
    eprint = "2510.01320",
    archivePrefix = "arXiv",
    primaryClass = "hep-ph",
    month = "10",
    year = "2025"
}

@article{Gan:2025nlu,
    author = "Gan, Xucheng and Liu, Da and Liu, Di and Luo, Xuheng and Yu, Bingrong",
    title = "{Detecting Ultralight Dark Matter with Matter Effect}",
    eprint = "2504.11522",
    archivePrefix = "arXiv",
    primaryClass = "hep-ph",
    reportNumber = "DESY-25-060",
    month = "4",
    year = "2025"
}

@article{Essig:2024wtj,
    author = "Essig, Rouven",
    title = "{Some progress {\&} challenges for the direct-detection of sub-GeV dark matter}",
    doi = "10.1016/j.nuclphysb.2024.116484",
    journal = "Nucl. Phys. B",
    volume = "1003",
    pages = "116484",
    year = "2024"
}

@article{Baker:2023kwz,
    author = "Baker, Christopher G. and Bowen, Warwick P. and Cox, Peter and Dolan, Matthew J. and Goryachev, Maxim and Harris, Glen",
    title = "{Optomechanical dark matter instrument for direct detection}",
    eprint = "2306.09726",
    archivePrefix = "arXiv",
    primaryClass = "hep-ph",
    doi = "10.1103/PhysRevD.110.043005",
    journal = "Phys. Rev. D",
    volume = "110",
    number = "4",
    pages = "043005",
    year = "2024"
}

@article{Du:2022dxf,
    author = "Du, Peizhi and Ega{\~n}a-Ugrinovic, Daniel and Essig, Rouven and Sholapurkar, Mukul",
    title = "{Doped semiconductor devices for sub-MeV dark matter detection}",
    eprint = "2212.04504",
    archivePrefix = "arXiv",
    primaryClass = "hep-ph",
    doi = "10.1103/PhysRevD.109.055009",
    journal = "Phys. Rev. D",
    volume = "109",
    number = "5",
    pages = "055009",
    year = "2024"
}

@inproceedings{Essig:2022dfa,
    author = "Essig, Rouven and others",
    title = "{Snowmass2021 Cosmic Frontier: The landscape of low-threshold dark matter direct detection in the next decade}",
    booktitle = "{Snowmass 2021}",
    eprint = "2203.08297",
    archivePrefix = "arXiv",
    primaryClass = "hep-ph",
    reportNumber = "FERMILAB-CONF-22-181-PPD",
    month = "3",
    year = "2022"
}

@article{Hochberg:2025dom,
    author = "Hochberg, Yonit and Novko, Dino and Ovadia, Rotem and Politano, Antonio",
    title = "{Unconventional Materials for Light Dark Matter Detection}",
    eprint = "2507.07164",
    archivePrefix = "arXiv",
    primaryClass = "hep-ph",
    month = "7",
    year = "2025"
}

@article{Helis:2024vhr,
    author = "Helis, D. L. and others",
    title = "{First measurement of Gallium Arsenide as a low-temperature calorimeter}",
    eprint = "2404.15741",
    archivePrefix = "arXiv",
    primaryClass = "hep-ex",
    doi = "10.1140/epjc/s10052-024-13123-8",
    journal = "Eur. Phys. J. C",
    volume = "84",
    number = "7",
    pages = "749",
    year = "2024"
}

@article{Das:2023cbv,
    author = "Das, Anirban and Jang, Jiho and Min, Hongki",
    title = "{Sub-MeV dark matter detection with bilayer graphene}",
    eprint = "2312.00866",
    archivePrefix = "arXiv",
    primaryClass = "hep-ph",
    doi = "10.1103/PhysRevD.110.043020",
    journal = "Phys. Rev. D",
    volume = "110",
    number = "4",
    pages = "043020",
    year = "2024"
}

\end{document}